\documentclass[aps,prb,twocolumn,showpacs]{revtex4}
\usepackage{amssymb}
\usepackage{graphicx}

\begin{document}

\title{Structural and transport properties of Sr$_2$VO$_{3-\delta}$FeAs superconductors with different oxygen deficiencies}

\author{Fei Han, Xiyu Zhu, Gang Mu, Peng Cheng, Bing Shen, Bin Zeng and Hai-Hu Wen}\email{hhwen@aphy.iphy.ac.cn }

\affiliation{National Laboratory for Superconductivity, Institute of
Physics and Beijing National Laboratory for Condensed Matter
Physics, Chinese Academy of Sciences, P. O. Box 603, Beijing 100190,
China}

\begin{abstract}
Sr$_2$VO$_{3-\delta}$FeAs superconductors with different oxygen
deficiencies have been successfully fabricated. It is found that the
superconducting transition temperature drops down monotonically with
the increase of oxygen deficiency. The diminishing of
superconductivity is accompanied by the enhancement of residual
resistivity, indicating an unraveled scattering effect induced by
the oxygen deficiency. The highest superconducting transition
temperature at about 40 K is achieved near the stoichiometrical
sample Sr$_2$VO$_{3}$FeAs. Surprisingly, the X-ray photoelectron
spectroscopy (XPS) shows that the vanadium has a "5+" valence state
in the samples. The Hall effect measurements reveal that the density
of charge carriers (electron-like here) varies qualitatively with
the increase of oxygen deficiency. Magnetotransport measurements
show that the superconducting transition changes from one-step-like
shape at low fields to two-step-like one at high fields, indicating
a high anisotropy.
\end{abstract}
\maketitle

\section{Introduction}

The iron-based superconductors have formed a new family in the field
of high-$T_c$ superconductivity.\cite{Kamihara2008,WenAdvMat2008}
Very soon, many new structures with the FeAs layers have been found,
including the so-called 1111 phase (LNFeAsO, AEFeAsF, LN = rare
earth elements, AE = alkaline earth
elements),\cite{Kamihara2008,Matsuishi2008,SrFeAsF} 122 phase
(AEFe$_2$As$_2$, AE = alkaline earth
elements),\cite{Johrendt,ChuCW2,Mandrus} 111 phase (LiFeAs,
NaFeAs),\cite{JinCQ,ChuCW} and 11 phase (FeSe).\cite{WuMK} Inspired
by the experience in the cuprates, a higher $T_c$ can be achieved in
a system with an expanded c-axis lattice, therefore, to explore new
systems with much expanded c-axis lattice constants is highly
desired. Towards this direction, the FeAs-based compound
Sr$_3$Sc$_2$O$_5$Fe$_2$As$_2$ with perovskite structure was found by
our group.\cite{FeAs32522} Unfortunately no superconductivity was
found in this compound. Later on, a new FeP-base compound
Sr$_2$ScO$_3$FeP with perovskite structure has been found and shown
to be a new superconductor at 17 K.\cite{ScFeP21311} Recently, we
successfully fabricated a new superconductor Sr$_2$VO$_{3}$FeAs with
$T_c$ = 37.2 K.\cite{VFeAs21311} By applying a high pressure to
Sr$_2$VO$_{3}$FeAs, the $T_c$ has been raised to 46
K.\cite{hpVFeAs21311} Several theoretical interests were raised on
this interesting superconductor. It was suggested that both Fe and V
contribute quasiparticle density of states (DOS) at the Fermi energy
$E_F$, while the electrons from the vanadium are 100\% spin
polarized.\cite{Shein} Also based on the band structure calculation,
Lee and Pickett concluded that the Fe-derivative orbitals do not
have the nesting condition,\cite{Pickett} therefore the model of
superconductivity mechanism based on the interpocket scattering of
electrons through exchanging anti-ferromagnetic spin fluctuations
was questioned. Recently it was argued that the nesting condition of
Fe-derivative bands may still hold in this material, although the
nesting condition is not as good as in other systems.\cite{Mazin}
Therefore in this system, there are several important questions to
be answered: (1) Whether the superconductivity is induced by the
oxygen deficiency or multi-valence state of vanadium? (2) Is the
system really a highly anisotropic one as expected from the
structure parameters? (3) Is there a nesting condition for the Fermi
surfaces of the Fe-derivative bands or not? In this paper, we report
the fabrication and characterization of the superconducting system
Sr$_2$VO$_{3-\delta}$FeAs with different oxygen contents.

\section{Experimental}

By using the solid state reaction method,\cite{ZhuXY} we
successfully fabricated the superconducting system
Sr$_2$VO$_{3-\delta}$FeAs with different oxygen contents. Firstly,
FeAs, and SrAs powders were obtained by the chemical reaction method
with Fe powders (purity 99.99\%), Sr pieces (purity 99.9\%) and As
grains (purity 99.999\%). Then they were mixed with V$_2$O$_{3}$
(purity 99.9\%), SrO (purity 99\%), and Fe powders (purity 99.99\%)
in the formula Sr$_2$VO$_{3-\delta}$FeAs, ground and pressed into a
pellet shape. All the weighing, mixing and pressing procedures were
performed in a glove box with a protective argon atmosphere (both
H$_2$O and O$_2$ are limited below 0.1 ppm). The pellet was sealed
in a silica tube under 0.2 atm argon atmosphere and followed by a
heat treatment at 1050 $^o$C for 30 hours. Then it was cooled down
slowly to room temperature. The X-ray diffraction (XRD) patterns of
our samples were carried out by a $Mac$-$Science$ MXP18A-HF
equipment with $\theta-2\theta$ scan. The XRD data taken on powder
samples was analyzed by the Rietveld fitting method using the GSAS
suite.\cite{GSAS} The DC susceptibility of the samples was measured
on a superconducting quantum interference device (SQUID, MPMS-7T) of
Quantum Design. The resistivity and Hall effect measurements were
done using a six-probe technique on the Quantum Design instrument
physical property measurement system with magnetic fields up to 9 T
(PPMS-9T). The temperature stabilization was better than 0.1\% and
the resolution of the voltmeter was better than 10 nV.

\section{Results and discussion}
\subsection{Superconductivity tuned by oxygen deficiency}

\begin{figure}
\includegraphics[width=9cm]{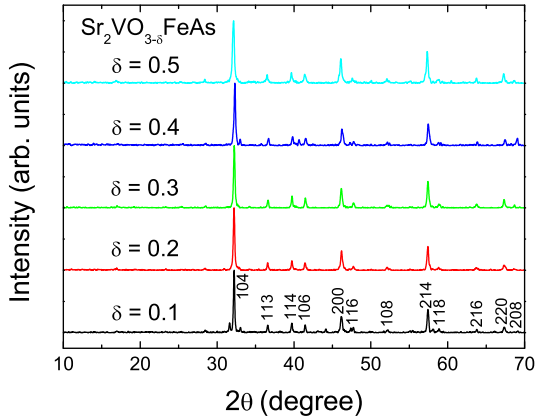}
\caption{(Color online) X-ray diffraction patterns for the samples
Sr$_2$VO$_{3-\delta}$FeAs. One can see that all the main peaks can
be indexed to the structure of FeAs-21311 with the space group of
\emph{P4/nmm}.} \label{fig1}
\end{figure}

In Fig. 1, we present the x-ray diffraction (XRD) patterns of the
compounds Sr$_2$VO$_{3-\delta}$FeAs with $\delta$ from 0.1 to 0.5.
These compounds contain a stacking of anti-fluorite (Fe$_2$As$_2$)
layers and perovskite-type (Sr$_4$V$_2$O$_6$) layers. For these
samples, all the main peaks can be indexed to the tetragonal
structure with the space group \emph{P4/nmm}. The impurity phase was
found to come from Sr$_2$VO$_4$. It is found that the lattice
constants change only slightly upon the change of oxygen content.
Taking Sr$_2$VO$_{2.5}$FeAs as an example, the lattice constants
were determined to be a = 3.927$\AA$ and c = 15.666$\AA$, while a =
3.928$\AA$ and c = 15.669$\AA$ for Sr$_2$VO$_{2.9}$FeAs. Comparing
the two samples, we may conclude that although the oxygen contents
of these samples change a lot, but the structure will relax by
itself and the lattice constants change slightly.

\begin{figure}
\includegraphics[width=9cm]{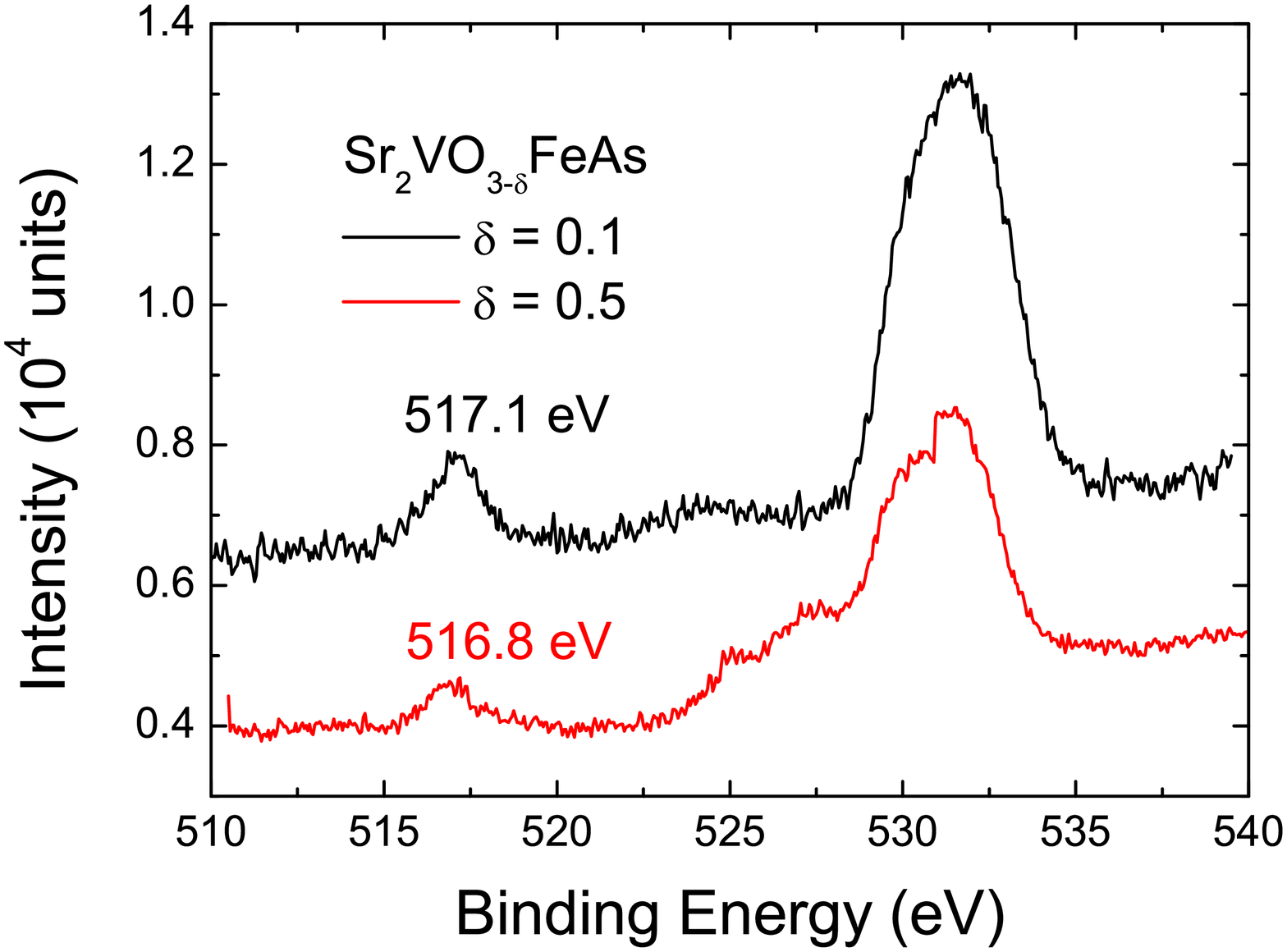}
\caption{(Color online) The X-ray photoelectron spectroscopy (XPS)
on Sr$_2$VO$_{3-\delta}$FeAs.} \label{fig2}
\end{figure}

As we know, the vanadium in the perovskite structure has multiple
valence states. For example, the vanadium has a valence state of
"3+" in the LaSrVO$_4$, while that in Sr$_2$VO$_4$ is "4+". In order
to determine the valence states of our samples, we measured the
X-ray photoelectron spectroscopy (XPS) of the samples
Sr$_2$VO$_{3-\delta}$FeAs with $\delta$ = 0.1, 0.5, as shown in Fig.
2. As we can see, there are peeks at about 517 eV which means the
vanadium may be on "5+" valence state in the samples. With reducing
oxygen content, the main peak of vanadium shifts slightly from 517.1
eV to 516.8 eV, suggesting that the vanadium has a quite stable
valence state. The V$^{5+}$ state is however unexpected by a simple
counting on the electrons. Assuming that the (FeAs) has a "-1"
valence state, and the cationic state Sr$^{2+}$ and O$^{2-}$, we
then have two electrons doped to each (FeAs). This is a highly doped
state compared to that in LaFeAsO$_{1-x}$F$_x$ and
Ba(Fe$_{1-x}$Co$_x$)$_2$As$_2$. However as we know the XPS detects
only the valence state on the surface layer, so we cannot conclude
definitely that the vanadium is at a valence state of "5+" in the
interior part of the samples. Future experiments are strongly
desired to resolve this puzzle. It may be safe to conclude that the
superconductivity achieved in Sr$_2$VO$_{3-\delta}$FeAs is because
the vanadium contributes large amount of electrons to the (FeAs)
planes.

\begin{figure}
\includegraphics[width=9cm]{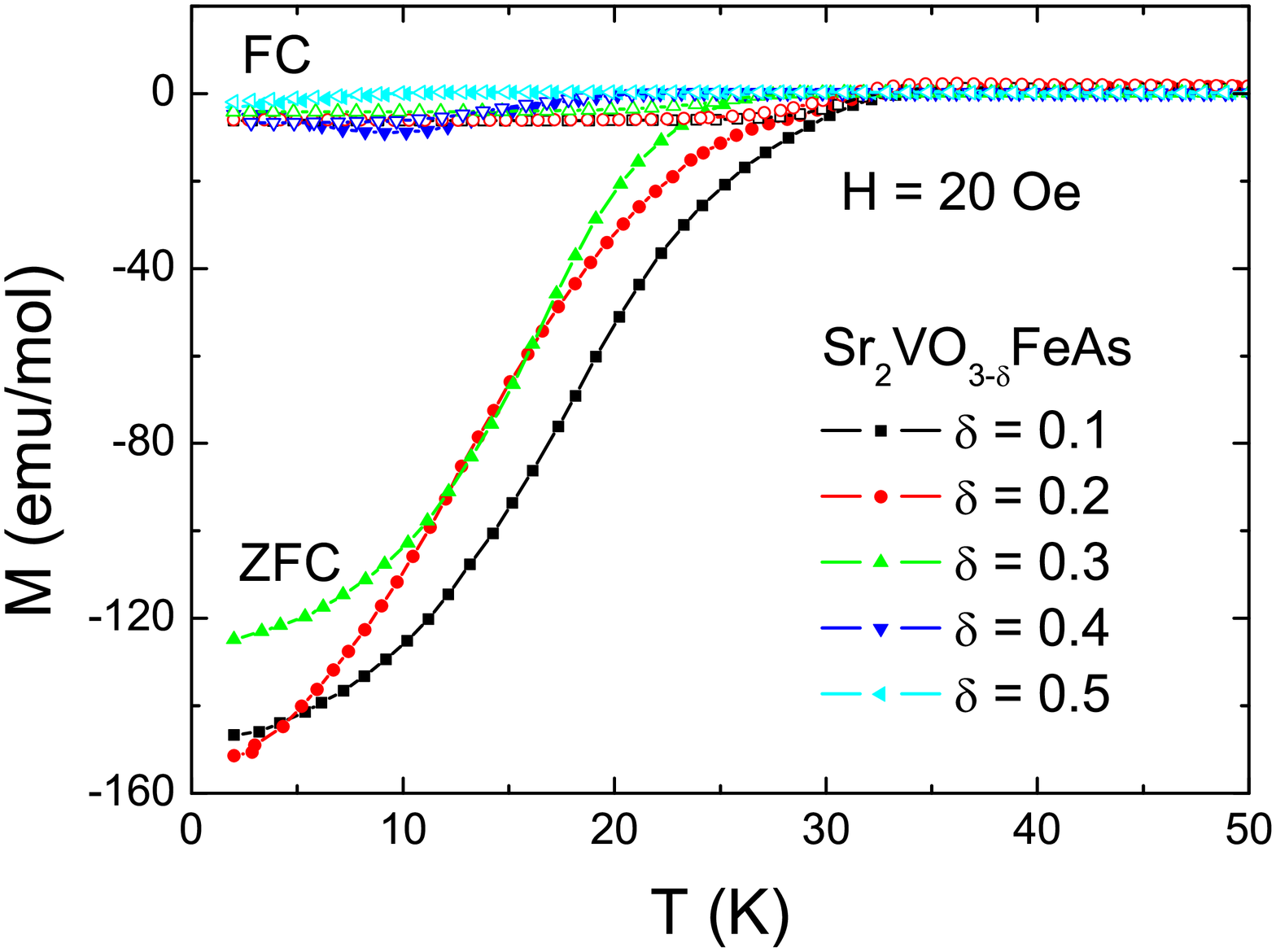}
\caption{(Color online) Temperature dependence of DC magnetization
for the samples Sr$_2$VO$_{3-\delta}$FeAs. The DC susceptibility
data was obtained using the zero-field-cooling and field-cooling
modes under a magnetic field of 20 Oe. For the oxygen deficient
samples with $\delta$ = 0.4, 0.5, the diamagnetization signal is
much weaker than the others.} \label{fig3}
\end{figure}

Although the lattice constants and the valence state of vanadium do
not exhibit a clear change with the oxygen deficiency, the
superconductivity property is however strongly influenced. In Fig.
3, we show the temperature dependence of DC magnetization for
Sr$_2$VO$_{3-\delta}$FeAs with $\delta$ from 0.1 to 0.5. The
measurements were carried out under a magnetic field of 20 Oe in
zero-field-cooled and field-cooled processes. As we can see, the
superconducting transition temperature drops with the increase of
oxygen deficiency. When $\delta$ is 0.1, 0,2, and 0.3, the
diamagnetization signal is quite big. While $\delta$ is up to 0.4,
0.5, the diamagnetization signal becomes very small. One possible
reason is that there is a very weak superfluid density when plenty
of the Cooper pairs are broken by the disorders induced by oxygen
deficiencies.

\begin{figure}
\includegraphics[width=9cm]{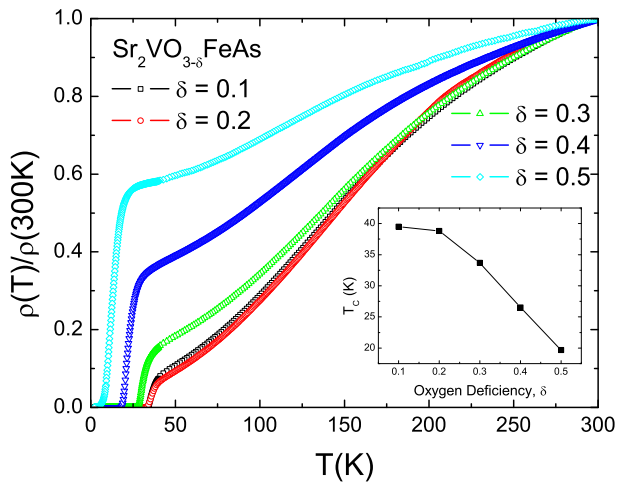}
\caption{(Color online) Electrical resistance of
Sr$_2$VO$_{3-\delta}$FeAs normalized to its value at 300K.The
transition temperature drops monotonically with the increase of
oxygen deficiency. The inset show the doping dependence of
superconducting transition temperature $T_c$. The highest
superconducting transition temperature is achieved near the
stoichiometric formula Sr$_2$VO$_{2.9}$FeAs. } \label{fig4}
\end{figure}

We also present the electrical resistance of the samples
Sr$_2$VO$_{3-\delta}$FeAs normalized to their values at 300 K in
Fig. 4. As we can see the superconducting transition temperature
drops down monotonically with the increase of oxygen deficiency. The
superconducting transition temperature was about 20 K for $\delta$ =
0.5, while it was about 40 K with $\delta$ = 0.1. It is clear that
there is an apparent change of behavior from a good metal to a bad
metal. As we can see the residual resistance ratio
$RRR\equiv\rho(300\;\mathrm{K})/\rho(30\;\mathrm{K})$ changes from
about 14 to about 1.7 when $\delta$ varies from 0.1 to 0.5. For the
sample with $\delta$ = 0.5, the residual resistivity becomes very
big. This may be understood that the electron conduction contributed
by the vanadium is strongly suppressed by the oxygen deficiency. An
alternative way to understand the large residual resistivity for
this oxygen deficient sample is that the oxygen deficiency leads to
an enhanced electrons scattering in the (FeAs) planes: The local
oxygen deficiency will twist the V-O semi-octahedron leading to a
severe influence on the local structure of the (FeAs) quasi 2D
planes. It remains unclear why the transition temperature drops
sharply with the increase of oxygen deficiency. We believe there is
a close relationship between the suppression of superconducting
transition temperature and the enhancement of residual resistivity.
We thus naturally conclude that the pair breaking effect caused by
disorders (as evidenced by the stronger residual resistivity) may be
the reason for the suppression of superconductivity. The inset of
Fig. 4 shows the doping dependence of the transition temperature
$T_c$. One can see that the highest $Tc$ is achieved at the
stoichiometric formula Sr$_2$VO$_{2.9}$FeAs. Therefore naturally we
conclude that the superconductivity in the present system is not
induced by the oxygen deficiency.

\subsection{Hall effect}

\begin{figure*}
\includegraphics[width=18cm]{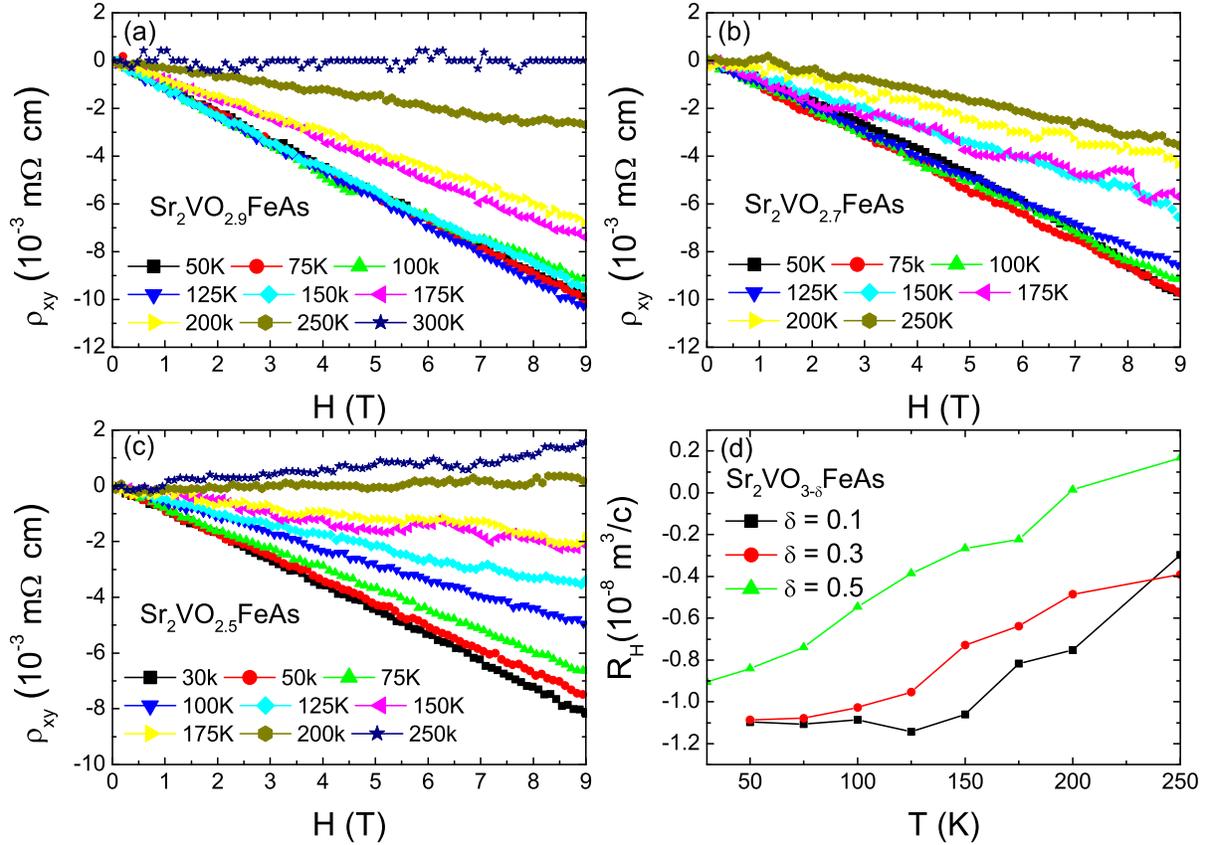}
\caption{(Color online) (a)-(c) The Hall resistivity $\rho_{xy}$
versus the magnetic field $\mu_0H$ at different temperatures for
Sr$_2$VO$_{3-\delta}$FeAs. (d) The absolute Hall coefficient drops
down with the increase of oxygen deficiency.} \label{fig5}
\end{figure*}

In order to investigate how the oxygen deficiency influences the
superconducting behavior, we measured the Hall effect in the normal
state with $\delta$ = 0.1, 0.3, 0.5. Fig. 5(a)-5(c) show the
magnetic field dependence of Hall resistivity ($\rho_{xy}$) at
different temperatures. In the experiment, $\rho_{xy}$ was taken as
$\rho_{xy} = [\rho(+H) - \rho(-H)]/2$ at each point to eliminate the
effect of the misaligned Hall electrodes. The raw data of the
transverse resistivity $\rho_{xy}$ are all negative and exhibits a
linear relation with the magnetic field. This is similar to that in
other FeAs-based superconductors.\cite{ZhuXY} Fig. 5(d) shows the
negative Hall coefficients $R_H = \rho_{xy}/H$ of the three samples.
As we can see, the absolute Hall coefficient drops with the increase
of oxygen deficiency, which means that the density of electron-like
charge carriers drops with the decrease of oxygen deficiency. The
strong temperature dependent behavior of the Hall coefficient $R_H$
suggests either a strong multi-band effect or a spin related
scattering effect. It is interesting to note that the sample with
higher $T_c$ has a weaker temperature dependence of $R_H$. This is
similar to that in other systems. For example, in
LaFeAsO$_{1-x}$F$_x$, much weaker temperature dependence is observed
for the optimally doped sample. While the $R_H$ of the underdoped
sample exhibits a very strong temperature dependence.

\subsection{High anisotropy of superconductivity}

\begin{figure*}
\includegraphics[width=18cm]{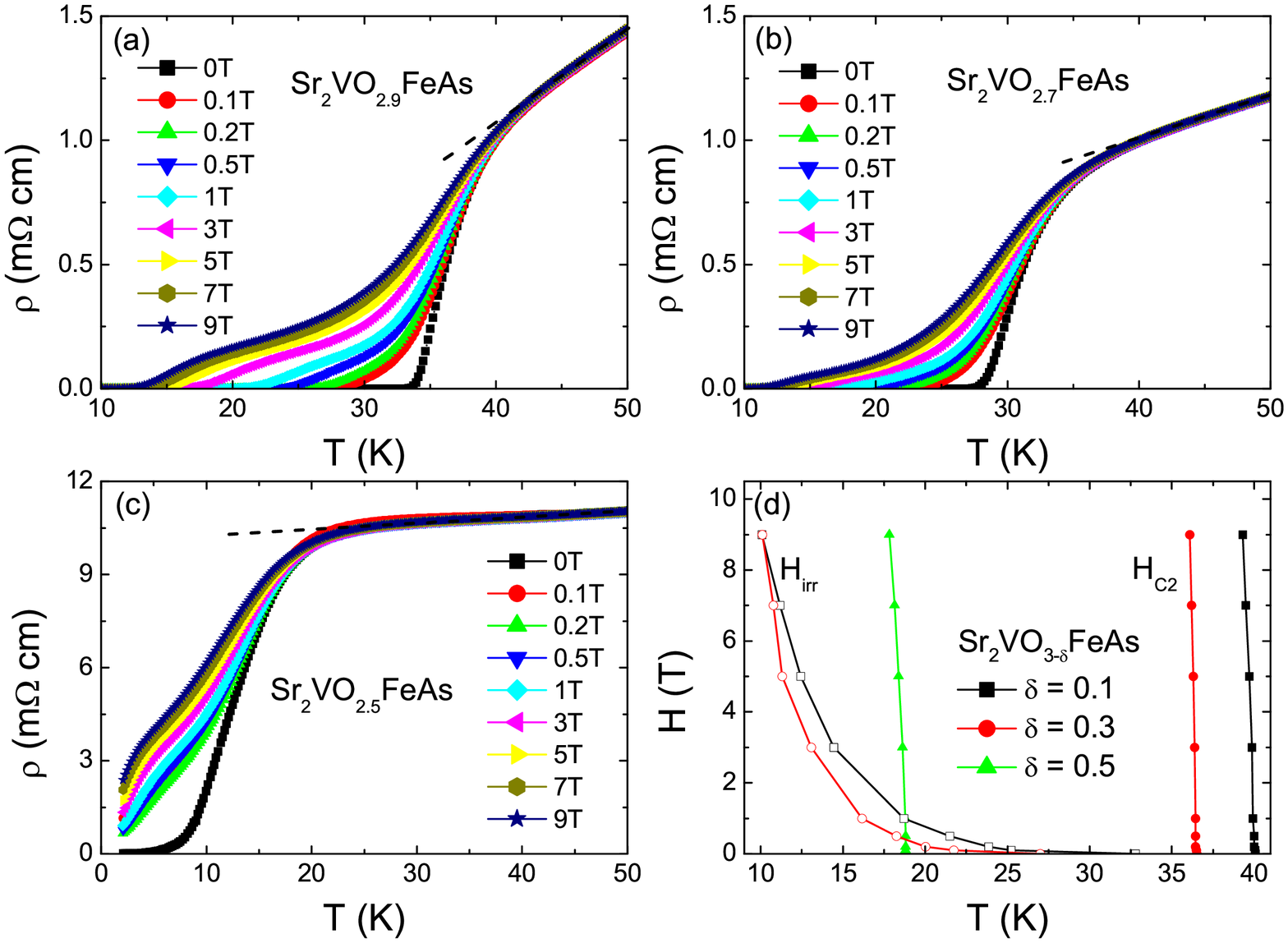}
\caption{(Color online) (a)-(c) Temperature dependence of
resistivity in the low temperature region under different magnetic
fields for Sr$_2$VO$_{3-\delta}$FeAs. (d) The phase diagram plotted
as H versus T. A criterion of 95\%$\rho_n$ was taken to determine
the upper critical fields, 0.1\%$\rho_n$ for the irreversibility
line $H_{irr}$. One can see that the one-step transition at a low
field will evolve into a two-step transition at a high field.}
\label{fig6}
\end{figure*}

In Fig. 6(a)-6(c) we present the temperature dependence of
resistivity for the samples Sr$_2$VO$_{3-\delta}$FeAs ($\delta$ =
0.1, 0.3, 0.5) under different magnetic fields. The onset transition
temperature of superconductivity is very robust against the magnetic
field, just like other iron-pnictide superconductors. As we can see,
the superconducting transition evolves from one-step-like at low
fields to two-step-like at high fields. The second step at a lower
temperature may be caused by the inter-layer coupling. This feature
has not been found in the other iron-pnictide superconductors.
Actually, it was often observed in the highly anisotropic systems in
cuprate superconductors like Bi-2212 and Bi-2223. So we suppose the
inter-layer coupling field may play an important role here. We used
the criterion of 95\%$\rho_n$ to determine the upper critical field
and show the data in Fig. 6(d), and taking the criterion of
0.1\%$\rho_n$ for the irreversibility line $H_{irr}$. Furthermore we
got $(dH_{c2}/dT)_{T_c} \approx$ -12.19 T/K for $\delta$ = 0.1,
$(dH_{c2}/dT)_{T_c} \approx$ -11.23 T/K for $\delta$ = 0.3, and
$(dH_{c2}/dT)_{T_c} \approx$ -9.09 T/K for $\delta$ = 0.5. These
values are rather large which indicates rather high upper critical
fields in these systems. In order to determine the upper critical
field in the low temperature region, we adopted the
Werthamer-Helfand-Hohenberg (WHH) formula $H_{c2} =
-0.69(dH_{c2}/dT)_{T_c}T_c$.\cite{WHH} Finally we get $H_{c2}(0)$ =
337 T for $\delta$ = 0.1, 283 T for $\delta$ = 0.3 and 119 T for
$\delta$ = 0.5. As we can see there are large regions between the
upper critical $H_{c2}(T)$ and the irreversibility field
$H_{irr}(T)$. This region may correspond to the vortex liquid region
dominated by the motion of pancake vortices. The global shape of
vortex phase suggests a high anisotropy of the system. An exact
evaluation on the anisotropy of this system would rely on the data
measured from single crystals, which is actually underway.

\section{Conclusions}

We have successfully fabricated the superconducting systems
Sr$_2$VO$_{3-\delta}$FeAs with different oxygen deficiencies. It is
found that the lattice constants and the valence state of vanadium
do not change with the oxygen deficiency, while the superconducting
transition temperature drops down dramatically accompanied by the
strong enhancement of the residual resistivity. The highest
superconducting transition temperature at about 40 K was achieved
near the stoichiometrical sample Sr$_2$VO$_{3}$FeAs, therefore we
conclude that the superconductivity in the present sample is not
induced by the oxygen deficiency and the multi-valence state.
Magnetotransport measurements lead to the determination of the
vortex phase diagram which resembles to that of highly anisotropic
system, such as Bi-2212 and Bi-2223. Hall effect measurements
clearly indicate a multi-band feature of the electron conduction.

\section{Acknowledgements}

This work was supported by the Natural Science Foundation of China,
the Ministry of Science and Technology of China (973 Projects
No.2006CB601000, No. 2006CB921802), and Chinese Academy of Sciences
(Project ITSNEM).

\end{document}